\documentclass[checkin,showpacs,aps,prl]{revtex4-1}
\newcommand{\bk}{{\bf k}}

\newcommand{\ub}{\mu_{\rm B}}
\newcommand{\bn}{\begin{enumerate}}
\newcommand{\en}{\end{enumerate}}
\newcommand{\ba}{\begin{eqnarray}}
\newcommand{\ea}{\end{eqnarray}}

\usepackage{graphicx,color}

\newcommand{\be}{\begin{equation}}
\newcommand{\ee}{\end{equation}}

\newcommand{\ra}{\rangle}
\newcommand{\et}{{\it et al. }}
\newcommand{\ete}{{\it et al.}}

\def\prl{{ Phys. Rev. Lett. }}

\def\prb{{ Phys. Rev. B }}

\begin{document}














\newcommand{\clr}{\color{black}}




\title{Magnetic spin moment reduction in photoexcited ferromagnets
  through exchange interaction quenching: Beyond the rigid band
  approximation}

\author{G. P. Zhang$^{1,*}$, M. S. Si$^{2}$, Y. H. Bai$^{3}$, and
  Thomas F. George$^{4}$ }

 \affiliation{$^{1}$Department of Physics, Indiana State University,
   Terre Haute, IN 47809, USA }

 \affiliation{$^{2}$Key Laboratory for Magnetism and
  Magnetic Materials of the Ministry of Education, Lanzhou University,
  Lanzhou 730000, China}

\affiliation{$^{3}$Office of Information Technology,
  Indiana State University, Terre Haute, IN 47809, USA }

\affiliation{$^{4}$Office of the Chancellor and Center for Nanoscience \\Departments of
  Chemistry \& Biochemistry and Physics \& Astronomy \\University of
  Missouri-St. Louis, St.  Louis, MO 63121, USA }

\date{\today}

\begin{abstract}
{The exchange interaction among electrons is one of the most
  fundamental quantum mechanical interactions in nature and underlies
  any magnetic phenomena from ferromagnetic ordering to magnetic
  storage.  The current technology is built upon a thermal or magnetic
  field, but a frontier is emerging to directly control magnetism
  using ultrashort laser pulses.  However, little is known about the
  fate of the exchange interaction.  Here we report unambiguously that
  photoexcitation is capable of quenching the exchange interaction in
  all three $3d$ ferromagnetic metals.  The entire process starts with
  a small number of photoexcited electrons which build up a new and
  self-destructive potential that collapses the system into a new
  state with a reduced exchange splitting. The spin moment reduction
  follows a Bloch-like law as $M_z(\Delta E)=M_z(0)(1-{\Delta
    E}/{\Delta E_0})^{\frac{1}{\beta}}$, where $\Delta E$ is the
  absorbed photon energy and $\beta$ is a scaling exponent. A good
  agreement is found between the experimental and our theoretical
  results.  Our findings may have a broader implication for dynamic
  electron correlation effects in laser-excited iron-based
  superconductors, iron borate, rare-earth orthoferrites, hematites
  and rare-earth transition metal alloys.
}
\end{abstract}




 \maketitle

\section{1. Introduction}

Ultrafast laser technology fuels unprecedented investigations in
physics, chemistry, material science and technology. Using a
femtosecond laser pulse to steer chemical reactions is the foundation
of femtochemistry (Nobel prize in chemistry in 1999)
\cite{femtochemistry}. This inspires the development of femtosecond
Raman \cite{wang} and 2D IR spectroscopy \cite{tucker}.  A strong and
ultrafast laser pulse can rip off and drive back electrons from
gaseous atoms to generate high order harmonic generations, with
emitted energy exceeding 200 eV and with time duration on the order of
several hundred attoseconds (1 as=$10^{-18}$ s), representing an era
of attophysics \cite{attophysics}.  Ultrafast dynamics and
fragmentation of $\rm C_{60}$ were investigated under intense laser
pulses \cite{lc}.  Ultrafast laser pulses can coherently control the
four-wave mixing signals in GaAs \cite{qvu}. Efforts in
superconductors started one decade ago, with an enormous success, for
some latest discoveries, see \cite{htc1,htc2,htc3,htc4}.  A strong
laser field can even induce a transient superconductivity above $\rm
T_c$ in YBa$_2$Cu$_3$O$_{7-\delta}$ \cite{kaiser}, and reveals the
competition between the pseudogap and superconducting states
\cite{coslovich}. An ultrafast laser allows one to investigate charge,
spin and lattice dynamics in complex materials.  Just within a week, a
flurry of three research papers \cite{gerber,huisman,mik} reported
photoinduced dynamics in three entirely different systems: lattice
dynamics in high-temperature iron pnictide superconductors
\cite{gerber}, exchange parameter modification in iron oxides
\cite{mik}, and orbital magnetism in multisublattice metallic magnets
\cite{huisman}.

Laser-induced ultrafast demagnetization represents a major
breakthrough in magnetism. Beaurepaire and his colleagues \cite{eric}
demonstrated that a femtosecond laser pulse can induce an ultrashort
demagnetization in fcc Ni within 1 ps.  This field, which is termed
femtomagnetism, is rapidly growing \cite{ourreview,kir}, with
nonthermal switching observed \cite{stanciu,mangin}, and motivated new
developments in table-top high harmonic probe in complex magnetic
materials at M-edge, which is normally only accessible using
synchrotron radiation \cite{stamm}.  Coherent ultrafast magnetism is
also discovered by Barthelemy and colleagues \cite{bart,bigot1}. A new
comprehensive review is presented at the first conference on ultrafast
magnetism \cite{umc2013}.

Despite the enormous progress experimentally, theoretical
understanding falls behind.  In superconductors, besides an early
attempt \cite{jim}, only one study \cite{mik} presented a theoretical
analysis, but it does not catch the initial excitation of electrons
and subsequent change in the spin exchange interaction \cite{jpcm14}.
In magnetic materials, Sandratskii and Mavropoulos \cite{sand} found
that the Elliott-Yafet mechanism \cite{koo1} plays an important role
in femtomagnetic properties of FeRh, which complements the
superdiffusive mechanism \cite{opp} and the laser-spin-orbit coupling
mechanism \cite{prl00,np09}.  One important feature of these prior
theoretical studies is that they do not allow band structures to
change. This rigid band approximation has already been proven
inadequate for simple $3d$ transition metals
\cite{sch1,kra,sk,mueller,peter}.

 For instance, our first-principles calculation shows that under the rigid
 band approximation the induced spin change is less than 1\%
 \cite{jap08,prb09,jap15}.  There are several reasons why the spin
 change is small.  Si \et \cite{mingsu} showed that due to the laser
 photon energy $\hbar \omega$, only those transitions whose transition
 energy $\Delta E $ matches $\hbar\omega$ can be strongly excited,
 while others are optically silent. This limits on the number of
 electrons that can be excited. Once the number of excited electrons
 is small, then the spin change is likely to be small.  Essert and
 Schneider \cite{sch1} further showed that even including the
 electron-phonon interaction and the electron-electron interaction
 \cite{kra}, the spin moment change is very small.  In 2013, Mueller
 and coworkers \cite{mueller} employed a simple model system but
 included a feedback from the charge change; they found a substantial
 spin reduction. Krieger \et \cite{peter} carried out the
 time-dependent density functional investigation and found that the
 spin reduction is comparable to the experimental one, although their
 laser fluences were about 2-3 orders of magnitude higher than
 experimental fluences.

Besides those initial theoretical efforts, no study on the exchange
interaction change during photoexcitation has been carried
out. Nevertheless, these studies point out a possible solution. It is
possible that the band structure relaxation and self-consistency are
essential to our current understanding of the demagnetization process
in ferromagnets. The importance of research along this direction
should not be under-estimated since it may have a broader implication
in magnetic excitations in high-temperature superconductors.
Ultrafast laser and x-ray technology has a unique capability to
separate the spin excitation and phonon excitation on different time
scales, and provides new insights into the nature of these elementary
excitations. For instance, Chuang \et \cite{shen} employed the
time-resolved resonant x-ray diffraction to follow the strongly
coupled spin and charge order parameters in stripe-ordered nickelate
crystals.  Smallwood \et \cite{htc5} showed that one can even track
the Cooper pairs dynamics by ultrafast angle-resolved
photoemission. These experimental findings are exciting.  A
theoretical investigation on the exchange interaction change during
the photoexcitation is much needed.

Here we report the first density functional study of the exchange
interaction quenching during laser excitation.  We first construct an
excited potential energy surface by promoting a small number of
electrons from the valence band to the conduction band.  Even though
the number of electrons actually excited is small, the excited-state
potential is quite different from the ground-state potential when the
excited state is a few eV above the Fermi level.  Then we
self-consistently solve the Kohn-Sham equation under this excited
potential.  This self-consistency triggers an avalanche on the entire
system and importantly affects those unexcited electrons that are
initially unexcited, so that the exchange splitting is sharply
reduced.  For all the three $3d$ ferromagnets, we observe a big
reduction of spin moment.  If we assume 12.5\% absorption efficiency
of photon energy into fcc Ni, we can reproduce the same amount of
change observed experimentally \cite{eric}. Our theory can reproduce
the entire range of experimental fluence-dependence of the spin moment
change in bcc Fe \cite{weber} quantitatively for the same absorption
efficiency.  This is very encouraging.  The key to our success is that
we allow the full relaxation of the electronic band structure under
the excited potential. We expect that our formalism will move us one
step closer to reveal the true mechanism of femtomagnetism, and this
may also present a reliable method to investigate the spin excitation
in high temperature iron-based superconductors and metallic magnets
for the spin switching.

This paper is arranged as follows. In Section 2, we present our ideas
and theoretical scheme. Section 3 is devoted to the results and
discussion on the demagnetization, band relaxation and exchange
splitting reduction.  We conclude our paper in Section 4.


\section{2. Theoretical formalism}

Calculating excited states is traditionally a hard problem. The
progress in this field is slow and very limited, in comparison with
the ground state calculation. There is no easy and simple solution in
sight. The enormous development in ultrafast laser technology
presents new opportunities to investigate the charge and spin dynamics
on the femtosecond time scale in multiple fronts from traditional
high temperature superconductors, graphene, magnetic materials and
layer structures, topological insulators and nanostructures, to name
a few. Our effort represents a theoretical  effort in this direction.

\begin{figure}

\includegraphics[angle=0,width=12cm]{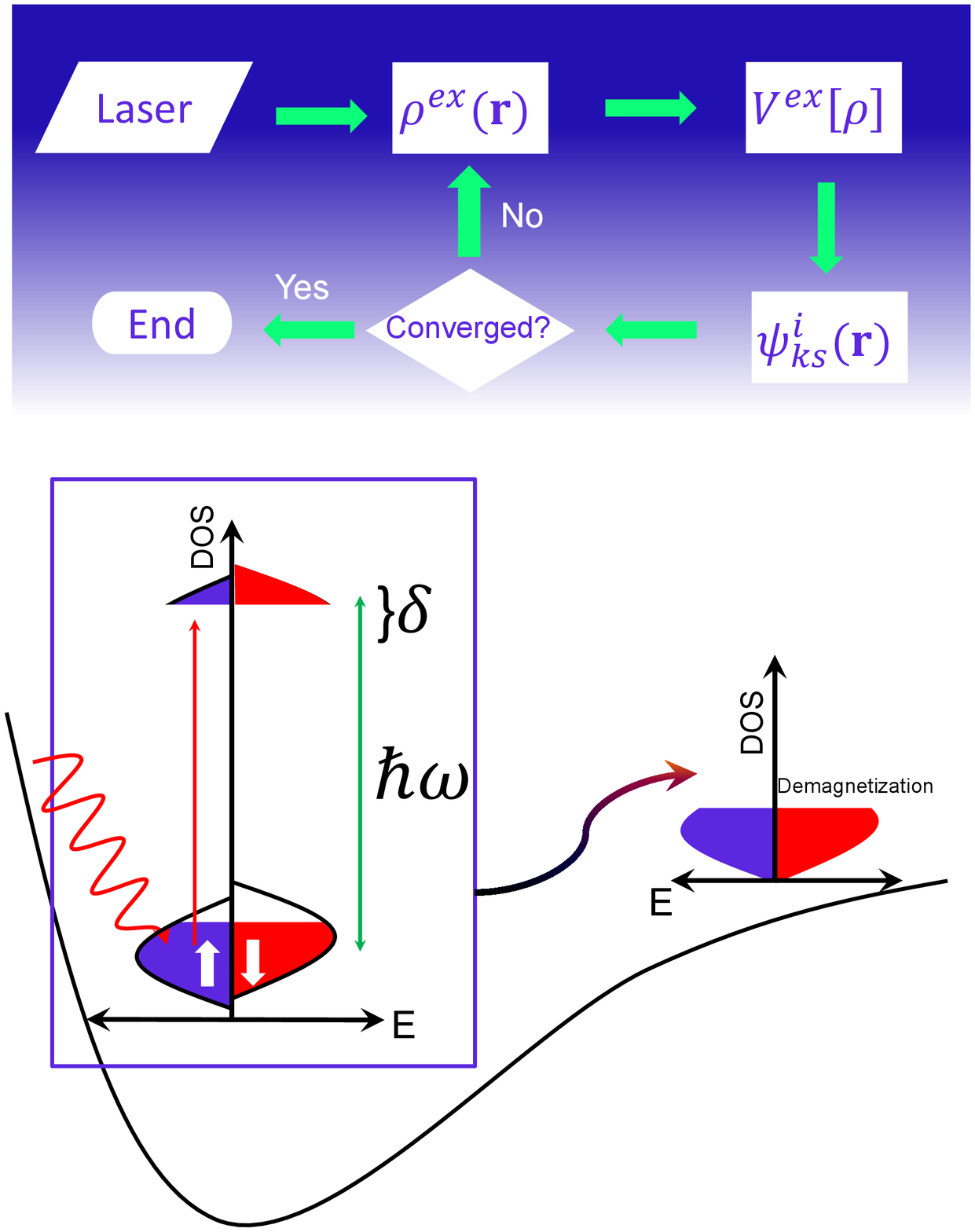}




\caption{ Strong demagnetization induced on an excited potential
  surface.  (Top) Computational scheme. The laser creates an excited
  charge density and excited potential energy surface for the entire
  system. By solving the Kohn-Sham (KS) equation, we attain the KS
  wavefunction for the next iteration until convergence.  (Bottom) The
  laser excites only a very small number of electrons out of the Fermi
  sea, but the generated potential affects all the electrons.  This
  drives the band structure relaxation, reduces the exchange
  splitting, and demagnetizes the sample. The laser excitation is
  determined by the photoenergy $\hbar\omega$, the width of the
  excitation $\delta$ (changing from 0.0 to 0.8 eV) and the strength
  of the excitation $\alpha$ (from 0.0 to 0.8, no unit).  }
\label{fig1}
\end{figure}

Figure \ref{fig1} schematically summarizes our main idea. When a laser
pulse impinges a magnet, it first promotes a few electrons from the
valence band $|\bk v\ra$ to the conduction band $|\bk c\ra$ (see the
bottom figure).  Due to energy conservation, the energy change $\Delta
E=E_{\bk c}-E_{\bk v} $ must be equal to the photon energy
$\hbar\omega$ of the laser within an energy window $\delta$ (inversely
proportional to the laser pulse duration).
 This initial excitation can
already induce some spin change \cite{prl00,np09,prb09}; and more
importantly, it directly affects the exchange interaction through \be
J(ab||ab)=\int\int d{\bf r}_1d{\bf r}_2 \phi_a^*({\bf r}_1)
\phi_b^*({\bf r}_2) \phi_a^*({\bf r}_2) \phi_b^*({\bf r}_1)|{\bf
  r}_1-{\bf r}_2|^{-1}, \ee where $\phi_{a(b)}({\bf r})$ is the
wavefunction, and the integration is over the electron coordinate
${\bf r}$.  For a free electron gas, with an increase in the kinetic
energy, the exchange energy decreases as \cite{mermin} \be
E_{ex}(k)=-\frac{2e^2}{\pi}k_f \left [\frac{1}2+\frac{1-x^2}{4x} \ln
  \left ( \frac{1+x}{1-x} \right )\right ], \ee where $x=k/k_f$ and
$k_f$ is the Fermi wavevector, and $k$ is the electron wavevector.

In the density functional theory, the exchange energy $E_{ex}[\rho]$
is a functional of the electron density $\rho({\bf r})$.  The effect
of the laser field enters through the excited density $\rho^{ex}({\bf
  r})=\sum_{{\bk}n}^{occ} n_{{\bk}n}= \sum_{{\bk}n}^{occ} | \psi_{{\bf
    k}n}({\bf r}) |^2$, which self-consistently generates a new
potential.  $\psi_{{\bf k}n}$ is the Kohn-Sham wavefunction computed
from \cite{sup}
\begin{widetext}
\be
\left [-\frac{\hbar^2}{2m}\nabla^2+v_{ext}({\bf r})+e^2\int \frac{\rho({\bf
    r}')}{|{\bf r}-{\bf r}'|} d{\bf r}'+v_{xc}[\rho({\bf r})] \right ]
\psi_{{\bf k}n}({\bf r}) =E_{{\bf k}n}\psi_{{\bf k}n}({\bf r}),
\ee
\end{widetext}
where the terms on the left-hand side are the kinetic energy, external
potential, Coulomb and exchange-correlation potential energies,
respectively. The spin-orbit coupling is included through the second
variational principle \cite{wien2k}.
$E_{{\bf k}n}$ and $\psi_{{\bf k}n}({\bf r})$ are the
eigenvalue and eigenwavefunction of state ${\bf k}n$.

 The top portion of Fig. \ref{fig1} shows the flow of our theoretical
 formalism.  For a pair excitation from $|\bk v\ra$ to $|\bk c\ra$, we
 construct the excited charge density via \cite{mingsu}
\ba
 \left  .  \begin{array}{ll}
n^{ex}_{\bk c}({\bf r}) &=\alpha n_{\bk v}({\bf r})+(1-\alpha)n_{\bk
  c}({\bf r})\\
 n^{ex}_{\bk v}({\bf r}) &=\alpha
n_{\bk c}({\bf r}) +(1-\alpha)n_{\bk v}({\bf r})\\
\end{array}
\right \} {\hspace{.5cm} \rm if} {\hspace{.5cm} |E_{\bk c}-E_{\bk
  v}-\hbar\omega|\le \delta}, \label{eq1}
\ea
 where $n_{\bk v} ({\bf r})$ and $n_{\bk c}({\bf r})$ are the charge
densities for the valence band ${\bk v}$ and conduction band ${\bk c}$,
respectively.
The weighted occupation of the
excitation, $\alpha$, represents the strength of the excitation and changes
from 0 to 1. If $\alpha=0$, this is just a ground-state calculation;
if the laser excitation is strong, $\alpha$ should be increased. If
the excitation energy falls outside $\delta$, no change is made to
their occupation. For this reason, $\delta$ should be kept reasonably
low, less than 1 eV; if it is too wide, characters of valence and
conduction bands may be quite different and vary a lot. This is
particularly important for  binary or ternary compounds.

Equation (\ref{eq1}) is missing in all the previous rigid-band
calculations. If the photoexcited valence and conduction bands had the
same orbital character, whether Eq. (\ref{eq1}) is included would not
make a big difference. But for the laser excitation with a few eV
above the Fermi level, the orbital characters of the valence and
conduction bands are quite different.  For this reason, we expect a
huge effect on the entire system.  Our method is similar to the
excitation energy calculation in transition metal atoms by Vukajlovic
\cite{vu} and rare-earth metals done by Herbst \et \cite{herbst}, and
more recently a photocarrier doping treatment \cite{wegkamp} (and also
quantum chemistry calculations). We implement our method using the
Wien2k code, which uses the full-potential augmented planewave
method. This code is among the most accurate density functional codes,
and is cheaper than other commercial codes, with open source codes and
well designed structures and directories (the reader is encouraged to
contact us for the further implementation details). One of the biggest
advantages over the pseudopotential codes is that it can be extended
to the core level excitation which has been a hot topic for the
experimental community.
In our supplementary materials, we provide all the details about our
implementation. Here, in brief, we summarize our major changes to the
code. The first major change is made to the {\tt lapw2}, where the new
charge density and potential are constructed through the above
equation \ref{eq1}. The second change is to add one input file which
includes the laser photon energy and energy window.
We revise the major scripts to run the code and also add
four new files which store the number of electron excited and the pair
indices of each excitation and their original weights.

\section{3. Results and Discussion}

\subsection{3.1 Demagnetization versus absorbed photon energy in Ni, Co
  and Fe }

\begin{figure}

\includegraphics[angle=270,width=14cm]{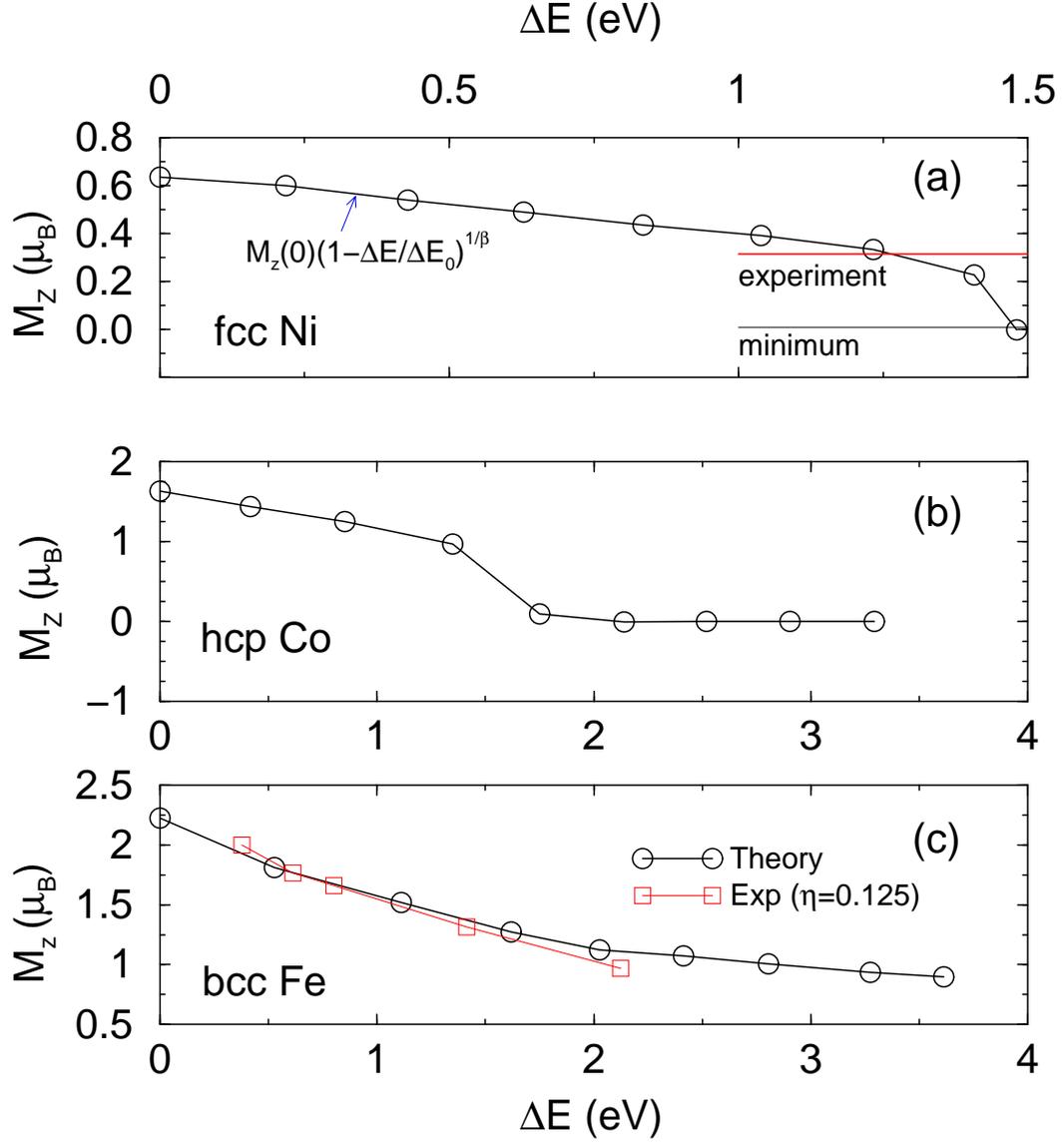}

\caption{Spin moment decreases as the energy absorbed increases for
  (a) fcc Ni, (b) hcp Co, and (c) bcc Fe. Here the photon energy is
  $\hbar\omega=2.0$ eV, and the excitation window is fixed at
  $\delta=0.5$ eV. For the same amount of energy absorbed by the
  system, fcc Ni is the easiest to be demagnetized, followed by hcp Co
  and bcc Fe, as expected from the strength of the magnetic
  ordering. In (a), a scaling function is shown.  The experimental
  spin reduction is highlighted by a red line. In (c), the empty boxes
  represent the experimental results from Weber \et \cite{weber},
  where the absorption efficiency factor $\eta$ is 0.125.  }
\label{fig2}
\end{figure}

Since the beginning of femtomagnetism, a central question is how the
spin moment reduction is correlated with the energy absorbed into a
system.  Figure \ref{fig2}(a) shows the spin moment in fcc Ni as a
function of the absorbed energy $\Delta E$ by changing $\alpha$, with
the excitation window fixed at $\delta=0.5$ eV.  Here $\Delta E$ is
defined as the total energy difference between the before-and-after
electron excitation, which is also called the promotion energy
\cite{vu}.  As the promotion energy increases, we find the spin moment
drops sharply.  This dependence can be fitted to a scaling that
resembles the magnetization curve, \be M_z(\Delta
E)=M_z(0)(1-\frac{\Delta E}{\Delta E_0})^{\frac{1}{\beta}}, \ee where
for fcc Ni, we find $M_z(0)=0.63\ub$, $\Delta E_0=1.48$ eV, and
$\beta=2.6$.  With this curve, in principle, we can compute the
average spin moment up to the penetration depth $d$ as \be \bar{M}_z=
\sum_{l=0}^L M_z (\Delta E_l)/L, \ee where $L$ is the number of atomic
layers up to the penetration depth, and $l$ is the layer
index. Unfortunately, the energy absorbed at each layer is unknown and
depends on the thickness of the sample, as shown by Schellekens \et
\cite{schellekens}, but no expression is given. We assume that the
absorbed energy is proportional to the light energy times an
absorption efficiency factor $\eta$, or $\Delta E_l= \frac{1}{2}\eta
E_{light} \exp(-la/2d) $, where $E_{light}$ is the light energy at the
top of the sample \cite{mingsu} and $a$ is the lattice constant.  For
any energy higher than $\Delta E_0$, the spin moment is zero.  At the
penetration depth, $\Delta E_l= \frac{1}{2}\eta E_{light}/e $.  The
red line in Fig. \ref{fig2}(a) denotes the experimental reduction
(50\%) \cite{eric}. We find that to have the same experimental spin
moment reduction at the penetration depth, $\eta=12.5\%$ is
enough. Obviously this $\eta$ is the most conservative estimate and
represents an uplimit since layers above the penetration depth must
have stronger demagnetization and by average the spin reduction is
larger than the experimental value.  This $\eta$ presents an
opportunity for the experimentalist to verify our theoretical
prediction.

We apply our theory to hcp Co (Fig. \ref{fig2}(b)) and bcc Fe
(Fig. \ref{fig2}(c)).  We see a similar spin reduction for hcp
Co. Once the absorbed energy is above 1.7 eV, the spin moment is
quenched completely. This critical energy is higher than in fcc Ni,
since hcp Co has a higher Curie temperature and is harder to be
demagnetized. The most difficult case is bcc Fe. To reduce its spin
moment by 50\%, one needs one photon per atom, which is consistent
with the experimental results.  Mathias \et \cite{mathias} found that
in the same experiment the Ni spin moment is quenched by 45\%, while
the Fe spin moment is quenched by only 19\%.

Up to now, all the comparisons between the experiments and our
theoretical results focus on a single laser fluence. Weber \et
\cite{weber} systematically investigated the dependence of spin moment
reduction in Fe on the pump pulse fluence. This presents an excellent
test case for our theory over a range of five pump fluences; and we
only have one tuning parameter $\eta$. It is important to point out
that tuning $\eta$ changes either the slope of the spin moment versus
the absorbed energy [$M_z-\Delta E$ curve in Fig. (\ref{fig2})] or the
absolute energy, but not both.  We use the same method as above and
find that only $\eta=12.5\%$ allows us to match both the absolute
energy absorbed and the slope of $M_z-\Delta E$ curve.  Figure
\ref{fig2}(c) shows that their experimental results (empty filled
boxes) agree our theoretical data (empty circles) within a few
percentage.  Such a quantitative agreement is encouraging as it gives
us more confidence in our first-principles methods. From the
comparison between bcc Fe and fcc Ni, we see that $\eta$ shows a weak
dependence on the material in question, but this may be due to the
similarity between bcc Fe and fcc Ni.  Additional testing and
investigation is necessary using other materials \cite{eric98,boeglin}.
To compare the theoretical and experimental results, we need the
energy absorbed for each layer, ideally starting from one monolayer,
grown on a transparent substrate so little photon energy is absorbed
into the substrate. To minimize the heating effect, we suggest to use
a shorter laser pulse and lower repetition rate. This also suppresses
the phonon contribution, and targets on the magnetic excitation alone.

\subsection{3.2 Band relaxation and exchange splitting reduction}

The exchange splitting reduction and transient band structures are
clearly observed experimentally in gadolinium and
terbium \cite{carley}. Teichmann and colleagues \cite{teichmann} found
that in Gd, the spin-down band moves down by 0.07 eV and the spin-up
band moves up by 0.2 eV; in Tb, the shifts are 0.16 eV for both spin
channels. These experimental results are consistent with an earlier
study in fcc Ni \cite{durr1}.

To reveal some crucial insights into the demagnetization, we employ
fcc Ni as an example and start with our ground-state calculation,
whose density of $d$-states (DOS) is shown in Fig. \ref{fig3}(a),
where the Fermi energy is at zero. The exchange splitting between the
majority and minority spin DOS maxima is 0.82 eV. Figure \ref{fig3}(b)
shows the DOS for the excited-state configuration, where $\alpha=0.7$
and $\delta=0.5$ eV. While the excited DOS shape does not change much,
the majority and minority bands are clearly shifted, with the larger
shift in the majority band by as much as 0.5 eV toward the Fermi
level.  The splitting is reduced to 0.24 eV, consistent with the
experimental findings \cite{durr1}.  The exchange splitting reduction
is a precursor to the spin moment decrease.

\subsection{3.3 Spin moment reduction as the excited potential
  surface relaxes}

We can reveal further details how the spin moment is reduced during
the self-consistent iterations. As an example, we use the same
$\alpha$ and $\delta$ as Fig. \ref{fig2}.  The laser photon energy is
also fixed at $\hbar\omega=2.0$ eV.  Figure \ref{fig3}(c) shows that
for the first two iterations the spin moment change is very small, by
about 0.02 $\ub$, or about 3\%. However, this is already far larger
than the spin moment change found in our rigid-band simulation
\cite{jap08,jap09}. This further confirms our earlier observation
\cite{mingsu} that even though the electrons are promoted to the
excited states, the spin moment change is very small in all the rigid
band calculations. The main reason is because the number of electrons
excited is only limited to those ${\bf k}$ points where the transition
energies match the photon energy. Electrons at other ${\bf k}$ points
have no contribution to the spin moment change.  After the third
iteration, the excited potential generated by those excited electrons
has a dramatic impact on the entire system; as a result, the spin
moment drops precipitously.  Figure \ref{fig2}(d) shows that the spin
gradually converges to 0.23 $\ub$, with a net reduction of 0.4 $\ub$,
or 63\%. This high percentage spin loss is consistent with the
experimental findings.

\subsection{3.4 Effects of the  excitation strength and excitation
  window}

\begin{figure}
\includegraphics[angle=270,width=13cm]{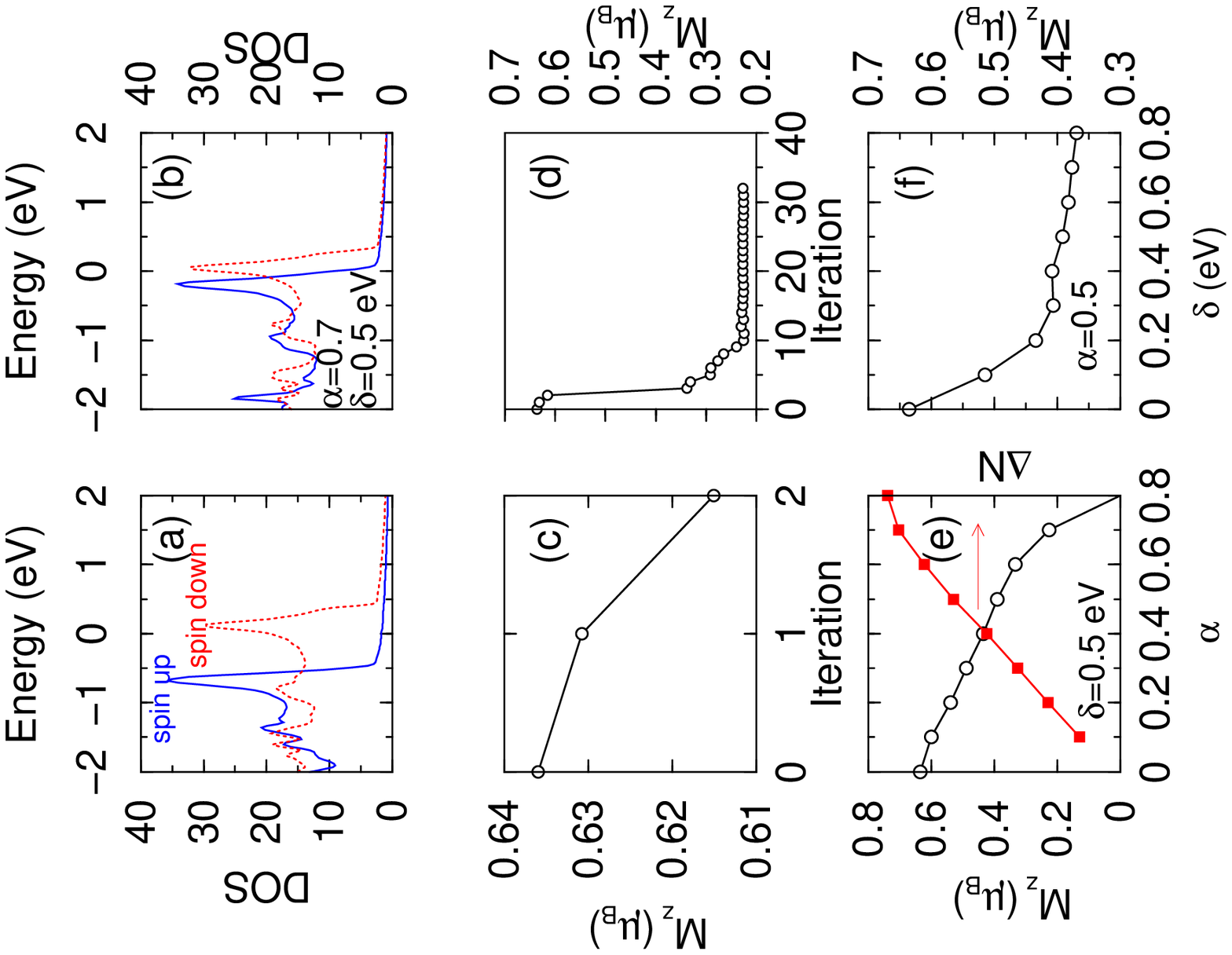}
\caption{ (a) Density of $d$-states in pristine fcc Ni.  The Fermi
  level is at 0 eV.  (b) Density of $d$-states in the excited state,
  with the exchange splitting clearly reduced.  Here the strength of
  excitation is $\alpha=0.7$, the laser photon energy is
  $\hbar\omega=2.0 $ eV, and the width of the excitation is
  $\delta=0.5$ eV.  (c) Small spin moment change for the first two
  iterations.  (d) Spin moment is sharply reduced as the
  self-consistent calculation iterates.  Iteration 0 refers to the
  converged case without excitation.  (e) Spin moment dependence
  (empty circles) on the excitation weight $\alpha$ for a fixed energy
  window ($\delta=0.5$ eV).  The filled boxes represent the number of
  electrons excited (Right axis).  (f) Spin moment reduction as a
  function of the excitation window $\delta$ for a fixed excitation
  weight at $\alpha=0.5$.  }
\label{fig3}
\end{figure}

To have a clear view as to how the level of excitation affects the
spin moment change, we keep the excitation energy window fixed and
gradually increase $\alpha$.  The empty circles in Fig.  \ref{fig3}(e)
show that as $\alpha$ increases from 0 to 0.8, the spin moment is
reduced precipitously and completely quenched at $\alpha=0.8$. We also
calculate the number of electrons actually excited. The filled boxes
in Fig.  \ref{fig3}(e) show how the number of excited electrons
changes with $\alpha$. In all the cases, the number of electrons
excited is below 1.  Quantitatively, we find that at $\alpha=0.1$, the
number of electrons excited is 0.13, and the spin reduction is 0.03
$\ub$, or 0.23 $\ub$ per electron. At $\alpha=0.7$, `0.7 electron' is
excited out of 10 valence electrons, and the spin is reduced by 0.4
$\ub$, so that for each electron excited, the spin is reduced by 0.57
$\ub$. This unambiguously demonstrates the importance of the
self-consistency and the band relaxation.

The excitation weight is not the only parameter that affects the spin
-- so does the excitation energy window $\delta$. Energetically, a
larger window corresponds to a shorter laser pulse.  With a larger
$\delta$, more states enter the excitation window. Figure
\ref{fig3}(f) shows that as the window becomes wider, a sharper
reduction is observed, but the change is not completely monotonic,
since the states moving in and out of the excitation window are not
continuous (see the density of states in Fig. \ref{fig3}(a)). We
should emphasize again that the value of $\delta$ should be relatively
small, less than 1 eV. In some cases, this also affects the
convergence (see the supplementary materials for details) \cite{sup}.

\section{4. Conclusions}

Through the first-principles density functional theory, we have
demonstrated unambiguously that even a small number of electrons
excited can lead to a strong quenching in the exchange interaction.
This process occurs through a band relaxation across the entire
Brillouin zone. The electrons in the excited states build a
self-destructive potential that greatly weakens the electron
correlation effect and reduces the exchange splitting. As a direct
consequence, the strong demagnetization is induced and the exchange
splitting is reduced, consistent with the experimental results
\cite{durr1}.  This resolves one of the most difficult puzzles in
femtomagnetism.  Our finding has a broader implication on the
ultrafast dynamics in iron pnictides since the laser can even change
the lattice structures \cite{gerber} and the exchange interaction must
be changed as well.  In iron oxides, the effect is even more important
\cite{mik}.  For the first time, our study establishes a different
paradigm: During the laser excitation it is the excitation of
electrons that impacts on the exchange interaction and spin moment,
while the effect of the spin fluctuation on the exchange interaction
is secondary and is on a much longer time scale.

\acknowledgments This work was solely supported by the U.S. Department
of Energy under Contract No. DE-FG02-06ER46304. Part of the work was
done on Indiana State University's quantum cluster and
high-performance computers.  The research used resources of the
National Energy Research Scientific Computing Center, which is
supported by the Office of Science of the U.S. Department of Energy
under Contract No. DE-AC02-05CH11231. This work was performed, in
part, at the Center for Integrated Nanotechnologies, an Office of
Science User Facility operated for the U.S. Department of Energy (DOE)
Office of Science by Los Alamos National Laboratory (Contract
DE-AC52-06NA25396) and Sandia National Laboratories (Contract
DE-AC04-94AL85000).

$^*$gpzhang@indstate.edu



\begin{thebibliography}{99}



\bibitem{femtochemistry}A. H. Zewail, Pure Appl. Chem. {\bf 72}, 2219
  (2000).

\bibitem{wang}W. Wang \ete, PNAS {\bf 110},  18397 (2013).

\bibitem{tucker}M. J. Tucker \ete, PNAS {\bf 110},  17314 (2013).
\bibitem{attophysics}P. H. Bucksbaum, Nature {\bf 421}, 593 (2002).



\bibitem{lc} Z. Z. Lin and X. Chen, Physics Letters A {\bf 377}, 797
  (2013)
\bibitem{qvu} Q. T. Vu \ete, \prl {\bf 89}, 3508 (2000).

\bibitem{htc1}R. A. Kaindl \ete,  Science {\bf 287}, 470 (2000).
\bibitem{htc2}R. D. Averitt \ete, Phys. Rev. B {\bf 63}, 140502 (2001).
\bibitem{htc3}N. Gedik \ete, Phys. Rev. Lett. {\bf 95}, 117005 (2005).
\bibitem{htc4}J. Demsar \ete,  Phys. Rev. Lett. {\bf 82}, 4918,
  (1999).

\bibitem{kaiser}S. Kaiser \ete, arxiv: 1205.4661 (2012).



\bibitem{coslovich} G. Coslovich \ete, \prl {\bf 110}, 107003 (2013).

\bibitem{gerber}S. Gerber, K. W. Kim, Y. Zhang, D. Zhu, N. Plonka,
  M. Yi, G. L. Dakovski, D. Leuenberger, P. S. Kirchmann, R. G. Moore,
  M. Chollet, J. M. Glownia, Y. Feng, J.-S. Lee, A. Mehta,
  A. F. Kemper, T. Wolf, Y.-D. Chuang, Z. Hussain, C.-C. Kao,
  B. Moritz, Z.-X. Shen, T. P. Devereaux, and W.-S. Lee,
  arXiv:1412.6842 (2014).

\bibitem{huisman} T. J. Huisman, R. V. Mikhaylovskiy, A. Tsukamoto,
  Th. Rasing, and A. V. Kimel, arXiv:1412.5396 (2014).




\bibitem{mik} R. V. Mikhaylovskiy, E. Hendry, A. Secchi,
  J. H. Mentink, M. Eckstein, A. Wu, R. V. Pisarev, V. V. Kruglyak,
  M. I. Katsnelson, Th. Rasing, and A. V. Kimel,  arXiv:1412.7094 (2014).


\bibitem{eric}E. Beaurepaire, J. -C. Merle, A. Daunois, and
  J.-Y. Bigot,
  {Phys. Rev. Lett.} {\bf 76}, 4250 (1996).


\bibitem{ourreview}G. P. Zhang, W. H\"ubner, E.  Beaurepaire, and
  J.-Y. Bigot, Topics Appl. Phys.  {\bf 83}, 245 (2002).

\bibitem{kir}A. Kirilyuk, A. V. Kimel, and Th. Rasing,
  Rev. Mod. Phys. {\bf 82}, 2731 (2010).


\bibitem{stanciu}C. D. Stanciu, F. Hansteen, A. V. Kimel, A. Kirilyuk,
  A. Tsukamoto, A. Itoh, and Th. Rasing, \prl {\bf 99}, 047601 (2007).


\bibitem{mangin}S. Mangin \ete, Nature Materials {\bf 13}, 286
  (2014).



\bibitem{stamm} C. Stamm, T. Kachel, N. Pontius, R. Mitzner, T. Quast,
  K. Holldack, S. Khan, C.  Lupulescu, E.F. Aziz, M. Wietstruk,
  H. A. D\"urr, W. Eberhardt, Nat. Mater. {\bf 6}, 740 (2007).

\bibitem{bart}M. Barthelemy, M. Sanches Piaia, M. Vomir, H. Vonesh,
  J.-Y. Bigot, {\it Ultrafast Magnetism I}, edited by J.-Y. Bigot
  \ete, Springer Proceeedings in Physics {\bf 159}, page 214 (2015).



\bibitem{bigot1}J.-Y. Bigot, M. Vomir, and E. Beaurepaire, Nature
  Phys. {\bf 5}, 515 (2009).

\bibitem{umc2013}J.Y. Bigot, W. H\"ubner, Th. Rasing, and
  R. Chantrell, {\it Ultrafastmagnetism I}, Springer Proceedings in
  Physics, Vol. 159 (2015)

\bibitem{jim}M. Sentef \ete, Phys. Rev. X {\bf 3}, 041033 (2013).



\bibitem{jpcm14}G. P. Zhang, M. Q. Gu and X. S. Wu, J. Phys.:
  Condens. Matter {\bf 26}, 376001 (2014).



\bibitem{sand} L. M. Sandratskii and P. Mavropoulos, \prb {\bf 83},
  174408 (2011).

\bibitem{koo1}B. Koopmans \ete, Nat. Mater.  {\bf 9}, 259 (2010).

\bibitem{opp} M. Battiato, K. Carva, and P. M. Oppeneer,
  Phys. Rev. Lett. {\bf 105}, 027203 (2010).


\bibitem{prl00} G. P. Zhang and W. H\"ubner,
 { Phys. Rev. Lett.} {\bf 85},
  3025 (2000).
\bibitem{np09}G. P. Zhang, W. H\"ubner, G. Lefkidis, Y. Bai, and
  T. F. George,
{ Nat. Phys.} {\bf 5}, 499
  (2009).


\bibitem{sch1}S. Essert and H. C. Schneider,
{ Phys. Rev. B} {\bf 84}, 224405 (2011).



\bibitem{kra} M.  Krau$\ss$, T.  Roth, S.  Alebrand, D.  Steil, M.
  Cinchetti, M.  Aeschlimann, and H.  C.  Schneider,
{ Phys. Rev. B} {\bf 80}, 180407(R) (2009).



\bibitem{sk} A. J. Schellekens and B. Koopmans, Phys. Rev. Lett. {\bf
  110}, 217204 (2013).



\bibitem{mueller}B. Y. Mueller, A. Baral, S. Vollmar, M. Cinchetti,
  M. Aeschlimann, H. C. Schneider, and B. Rethfeld,
{ Phys. Rev. Lett.} {\bf 111}, 167204 (2013).

\bibitem{peter}K. Krieger, J. K. Dewhurst, P. Elliott,
  S. Sharma, and E. K. U. Gross,
arXiv: 1406.6607 (2014).

\bibitem{jap08} G. P. Zhang, Y. Bai, W. H\"ubner, G. Lefkidis, and
  T. F. George,
{ J. Appl. Phys.} {\bf 103}, 07B113 (2008).

\bibitem{prb09} G. P. Zhang, Y. Bai, and T. F. George,
{ Phys. Rev. B} {\bf 80}, 214415 (2009).

\bibitem{jap15} G. P. Zhang, M. S. Si, and T. F. George, Journal of
  Applied Physics {\bf 117}, 17D706 (2015).



\bibitem{mingsu}M. S. Si and G. P. Zhang,
{ AIP Advances} {\bf 2}, 012158 (2012).


\bibitem{shen}Y. D. Chuang, W. S. Lee, Y. F. Kung, A. P. Sorini,
  B. Moritz, R. G. Moore, L. Patthey, M. Trigo, D.-H. Lu,
  P. S. Kirchmann, M. Yi, O. Krupin, M. Langner, Y. Zhu, S. Y. Zhou,
  D. A. Reis, N. Huse, J. S. Robinson, R. A. Kaindl, R. W. Schoenlein,
  S. L. Johnson, M. Forst, D. Doering, P. Denes, W. F. Schlotter,
  J. J. Turner, T. Sasagawa, Z. Hussain, Z.-X. Shen, T. P. Devereaux,
  \prl {\bf 110}, 127404 (2013).


\bibitem{htc5} C. L. Smallwood \ete, Science {\bf 336}, 1137 (2012).

\bibitem{weber}A. Weber, F. Pressacco, S. G\"unther, E. Mancini,
P. M. Oppeneer, and C. H. Back, \prb {\bf 84}, 132412 (2011).



\bibitem{mermin}N. W. Ashcroft and N. D. Mermin, {\it Solid State
  Physics} (Harcourt, 1976).

\bibitem{sup} See the supplementary materials for the implementation
  details and the underlying rationale for our scheme.


\bibitem{wien2k} P. Blaha, K. Schwarz, G. K. H. Madsen, D. Kvasnicka,
  and J. Luitz. WIEN2k: An augmented plane wave + local orbitals
  program for calculating crystal properties (Karlheinz Schwarz,
  Techn. Universit\"at Wien, Austria, 2001).

\bibitem{vu}F. R. Vukajlovic, E. L. Shirley, and R. M. Martin,
{ Phys. Rev. B} {\bf 43},
  3994 (1991).
\bibitem{herbst}J. F. Herbst, R. E. Watson, and J. W. Wilkins.
{ Phys. Rev. B} {\bf 17}, 3089 (1978).

\bibitem{wegkamp}D. Wegkamp, M. Herzog, L.  Xian, M.  Gatti, P.
  Cudazzo, C.  L. McGahan, R.  E. Marvel, R.  F. Haglund, A.
  Rubio, M.     Wolf \ete,
  arXiv: 1408.3209 (2014).

\bibitem{schellekens} A. J. Schellekens, W. Verhoeven, T. N. Vader, and
  B. Koopmans,
{ Appl. Phys. Lett.} {\bf 102}, 252408 (2013).

\bibitem{mathias} S. Mathias \ete,
{ PNAS} {\bf 109}, 4792
  (2012).





\bibitem{eric98} E. Beaurepaire, M. Maret, V. Halte, J.-C. Merle,
  A. Daunois, and J.-Y. Bigot,
  { Phys. Rev. B} {\bf 58}, 12134 (1998).

\bibitem{boeglin}C. Boeglin, E. Beaurepaire, V. Halte,
  V. Lopez-Flores, C. Stamm, N. Pontius, H. A. D\"urr, and
  J.-Y. Bigot,
{ Nature} {\bf 465}, 458 (2010).

\bibitem{carley} R. Carley \ete, \prl {\bf 109}, 057401 (2012).
\bibitem{teichmann} M. Teichmann \ete, \prb {\bf 91}, 014425 (2015).

\bibitem{durr1} H. S. Rhie, H. A. D\"urr, and W. Eberhardt,
{ Phys. Rev. Lett.}
  {\bf 90}, 247201 (2003).



\bibitem{jap09}T. Hartenstein, G. Lefkidis, W. H\"ubner, G. P. Zhang,
  and Yihua Bai,
{ J. Appl. Phys.} {\bf 105}, 07D305
  (2009).

\end{thebibliography}
\end{document}